\documentclass[twocolumn,prb,showpacs,superscriptaddress]{revtex4}

\usepackage{psfrag}
\usepackage{graphicx}
\usepackage{amsmath}
\usepackage{amssymb}

\begin{document}
\title{Spatially resolved quantum plasmon modes in metallic nano-films from first principles}
\author{Kirsten Andersen}
\affiliation{Center for Atomic-scale Materials Design (CAMD),
  Department of Physics, Technical University of Denmark,DK - 2800
  Kgs. Lyngby, Denmark.}
\author{Karsten W. Jacobsen}
\affiliation{Center for Atomic-scale Materials Design (CAMD),
  Department of Physics, Technical University of Denmark,DK - 2800
  Kgs. Lyngby, Denmark.}
\author{Kristian S. Thygesen}
\email{thygesen@fysik.dtu.dk}
\affiliation{Center for Atomic-scale Materials Design (CAMD) and Center for Nanostructured Graphene (CNG),
  Department of Physics, Technical University of Denmark,DK - 2800
  Kgs. Lyngby, Denmark.}

\date{\today}

\begin{abstract}
  Electron energy loss spectroscopy (EELS) can be used to probe
  plasmon excitations in nanostructured materials with atomic-scale
  spatial resolution. For structures smaller than a few nanometers
  quantum effects are expected to be important, limiting the validity
  of widely used semi-classical response models. Here we present a
  method to identify and compute spatially resolved plasmon modes from
  first principles based on a spectral analysis of the dynamical
  dielectric function. As an example we calculate the plasmon modes of
  0.5-4 nm thick Na films and find that they can be classified as
  (conventional) surface modes, sub-surface modes, and a discrete set
  of bulk modes resembling standing waves across the film. We find
  clear effects of both quantum confinement and non-local response. The quantum plasmon modes
  provide an intuitive picture of collective excitations of confined
  electron systems and offer a clear interpretation of spatially
  resolved EELS spectra.
\end{abstract}

\maketitle 

\section{Introduction}
The dielectric properties of a material are to a large extent governed by the collective eigenmodes of its electron system known as plasmons\cite{1}. Advances in spectroscopy and materials preparation have recently made it possible to study and control light-matter interactions at the nanometer length scale where particularly the surface plasmons play a key role\cite{2,3}. While the ultimate goal of nanoplasmonics as a platform for ultrafast and compact information processing remains a challenge for the future, the unique plasmonic properties of metallic nanostructures have already been utilized in a number of applications including molecular sensors\cite{4,5}, photo-catalysis\cite{6} and thin-film solar cells\cite{7}. 
   
Electron energy loss spectroscopy (EELS) has been widely used to probe plasmon excitations in bulk materials and their surfaces. More recently, the use of highly confined electron beams available in transmission electron microscopes has made it possible to measure the loss spectrum of low-dimensional structures on a sub-nanometer length scale and with $<0.1$ eV energy resolution\cite{8}. Because the loss spectrum is dominated by the plasmons this technique provides a means for visualising the real-space structure plasmon excitations\cite{9}.

All information about the plasmons of a given material is contained in its frequency-dependent dielectric function, $\epsilon$, which relates the total potential in the material to the externally applied potential to linear order,
\begin{equation}\label{eq1}
\phi_{ext}(r,\omega)=\int \epsilon(r,r',\omega)\phi_{tot}(r',\omega)dr'
\end{equation}
 (We have specialized to the case of longitudinal fields which can be represented by scalar potentials). In the widely used Drude model, one neglects the spatial variation of $\epsilon$ and describes the frequency dependence by a single parameter, the bulk plasmon frequency, $\omega_p=\sqrt{ne^2/m\epsilon_0}$ where $n$ is the average electron density of the material\cite{2}. For metal surfaces the Drude model predicts the existence of surface plasmons with frequency $\omega_s=\omega_p / \sqrt{2}$. While the Drude model provides a reasonable description of plasmons in simple metal structures in the mesoscopic size regime it fails to account for the dispersion (wave vector dependence) of the plasmon energy in extended systems and predicts a divergent field enhancement at sharp edges. These unphysical results are to some extent remedied by the semi-classical hydrodynamic models which can account for spatial non-locality of $\epsilon$ and smooth charge density profiles at the metal-vacuum interface\cite{2,18,19}. Still, for materials other than the simple metals and for truly nanometer sized structures, predictive modeling of plasmonic properties must be based on a full quantum mechanical description\cite{20,21}. The calculation of plasmon energies and EELS spectra of periodic solids from first principles is a well established discipline of computational condensed matter physics\cite{22,23,24,25}. However, the application of these powerful methods to systematically explore the real space structure of plasmon excitations at the nanoscale has, to our knowledge, not been achieved previously.

\begin{figure*}
\begin{center}
\includegraphics[width=1.0\linewidth]{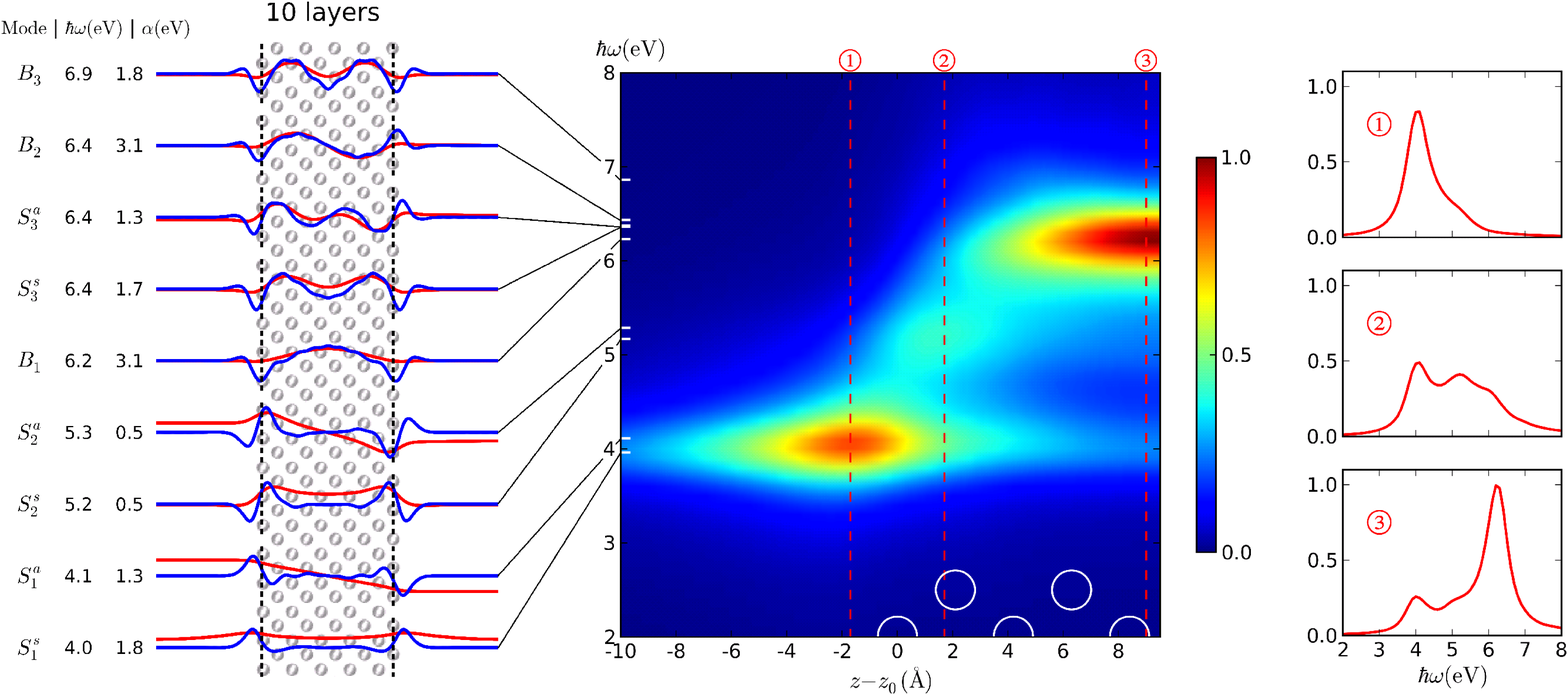}
\caption[system]{\label{fig1} (Color online)  Left: Spatial profile of the plasmon modes in the direction normal to the film for in-plane momentum transfer $q_\parallel$ = 0.1\AA$^{-1}$ . The red (blue) curves show the potential (density) associated with the plasmon excitation. Mode type, energy, and strength ($\alpha$) is shown to the left. Middle: Contour plot of the energy loss of an in-plane electron beam as it is scanned across the film. Only the left half of the film is shown and the position of the Na atoms is indicated by white circles at the bottom. The three bright features in the spectrum originate from energy loss due to excitation of three types of plasmons, namely the surface plasmon mode S$_1$, the subsurface mode S$_2$, and the lowest bulk plasmon modes B$_1$, respectively. Right: Calculated electron energy loss spectra along the dashed lines indicated in the middle panel.  }
\end{center}
\end{figure*}

In this paper we present a general method to calculate spatially resolved plasmon modes from first-principles. We apply the method to Na films of a few nanometer thickness where effects of quantum confinement and nonlocal response are expected to be important. The plasmon modes we find can be categorized as surface modes located mainly outside the metal surface, subsurface modes located just below the surface, and bulk modes. Very intuitively, the bulk plasmon modes resemble standing waves with nodes at the film surfaces. However, only plasmons with oscillation periods larger than~10 \AA~are found as a result of Landau damping which suppresses the strength of plasmon modes with smaller oscillation periods. Finally, we calculate the spatially resolved EELS spectrum of the metal films and show that all its features can be traced to excitation of specific plasmon modes.

\section{Method}
According to Eq. (\ref{eq1}), a self-sustained charge density oscillation, $\rho(r,\omega)$, can exist in a material if the related potential, satisfying the Poisson equation $\nabla^2 \phi = -4\pi \rho$  (atomic units are used throughout), obeys the equation
\begin{equation}
 \int \epsilon(r,r',\omega)\phi(r',\omega)dr' = 0
\end{equation}
In general, this equation cannot be exactly satisfied because the dielectric function will have a finite imaginary part originating from single-particle transitions which will lead to damping of the charge oscillation\footnote{We note that exact solutions may be obtained for complex frequencies. However, this requires an analytical continuation of $\epsilon$ into the complex plane and will not be pursued in the present work}. We therefore require only that the real part of $\epsilon$ vanishes and use the following defining equation for the potential associated with a plasmon mode,
\begin{equation}\label{eq.2}
 \int \epsilon(r,r',\omega_n)\phi_n(r',\omega_n)dr' = i\Gamma_n \phi_n(r,\omega_n)
\end{equation}
where $\Gamma_n$ is a real number. Mathematically, the plasmon modes are thus the eigenfunctions corresponding to purely imaginary eigenvalues of the dielectric function. Physically, they represent the potential associated with self-sustained charge-density oscillations damped by electron-hole pair formations at the rate $\Gamma_n$. 

It is instructive to consider the dielectric function in its spectral representation
\begin{equation}\label{eq.spec}
\epsilon(r,r',\omega)=\sum_n \epsilon_n(\omega)\phi_n(r,\omega)\rho_n(r',\omega),
\end{equation}
where the left and right eigenfunctions satisfy the usual orthonormality relation 
\begin{equation}\label{eq.ortho}
\langle \phi_n(\omega)|\rho_m(\omega)\rangle=\delta_{nm}.
\end{equation}
In the appendix we show that the left and right eigenfunctions are related through Poisson's equation
\begin{equation}\label{eq.dual}
\nabla^2 \phi_n(r,\omega) = -4\pi \rho_n(r,\omega),
\end{equation}
and thus correspond to the potential and charge density of the
dielectric eigenmode, respectively. Physically, the condition (\ref{eq.ortho}) expresses the fact that the different dielectric eigenmodes for a given frequency $\omega$, are electrodynamically decoupled.
The inverse dielectric function, $\epsilon^{-1}(r,r',\omega)$, is obtained by replacing the eigenvalues $\epsilon_n(\omega)$ in Eq. (\ref{eq.spec}) by $1/\epsilon_n(\omega)$. 

When the imaginary part of the eigenvalue $\epsilon_n(\omega)$ does not vary too much around the plasmon frequency $\omega_n$, the condition (\ref{eq.2}) is equivalent to the condition that
\begin{equation}\label{eq.max}
\text{Im}\epsilon_n(\omega_n)^{-1}\text{ is a local maximum.} 
\end{equation}
This is the case for most of the plasmon modes of the simple metal
films studied in this work. However, for the high energy bulk modes we found that the variation in $\text{Im}\epsilon_n(\omega)$
can shift the local maximum of $\text{Im}\epsilon_n(\omega)^{-1}$ away
from the point where $\text{Re}\epsilon_n(\omega)=0$. (The effect can
be even stronger of more complex materials with interband transitions.)
In such cases the condition (\ref{eq.max}) rather than (\ref{eq.2})
should be used to define the plasmon energy. Importantly, however, the
eigen functions $\phi_n(r,\omega)$ do not change
significantly when $\omega$ is varied between the frequencies given by
the two criteria and thus the spatial form of the plasmon remains well
defined.

In the case of metallic films where the in-plane variation of $\epsilon$ and $\phi_n$ can be assumed to have the forms $\exp(iq_\parallel\cdot (r_\parallel-r_\parallel '))$  and $\exp(iq_\parallel\cdot r_\parallel)$, respectively, Eq. (\ref{eq.2}) can be written
\begin{equation}
\int \epsilon(q_{\parallel},z,z',\omega_n)\phi_n(z',\omega_n)dz' = i\Gamma_n \phi_n(z,\omega_n)
\end{equation}
We stress that despite the atomic variation in the potential we have found that local field effects are negligible within the plane justifying this assumption. 

All the calculations were performed with the electronic structure code
GPAW\cite{28,29} using the Atomic Simulation Environment\cite{30}. The
Na films were modeled in a supercell with periodic boundary conditions
imposed in all directions. The minimal Na unit cell was used in the
plane of the film and 30 \AA~of vacuum was included in the direction
perpendicular to the film to separate the periodic images.  The
single-particle wave functions and energies were computed on a real
space grid with a grid spacing 0.18 \AA~and the Brillouin zone was
sampled by 64x64 $k$-points. The LDA functional was used for exchange
and correlation. The non-interacting density response function,
$\chi^0$, was calculated from the DFT single-particle states using a
50 eV energy cutoff for the plane wave basis and including states up
to 15 eV above the Fermi level. The frequency-dependence of the
dielectric function was sampled on the real frequency axis using a
grid from 0 to 20 eV with a 0.1 eV spacing. We checked that our
results were converged with respect to all the parameters of the
calculation. The microscopic dielectric function was calculated from
$\chi^0$ using the random phase approximation (RPA).  We note in
passing that this DFT-RPA level of theory yields highly accurate bulk
and surface plasmon energies (within a few tenths on an electron volt)
of the simple metals\cite{25}.

To obtain the plasmon modes, the dielectric function is diagonalized
in a plane-wave basis on each point of a uniform frequency grid. To
simulate an isolated film, the obtained eigenmode potentials were
corrected to remove the effect of interactions with films in the other
supercells, i.e. the periodic boundary conditions from the DFT
calculation (this is most important for small $q_{\parallel}$
vectors). In general the eigenmodes were found to be complex valued
reflecting the spatial variation in the phase of the plasmon
potential. This phase variation is, however, rather weak for the
present system, and by a suitable choice of the overall phase the
modes can be made almost real $(>90\%)$. For this reason only the real
part is shown in Figs. 1 and 3.

\section{Results}
\begin{figure}
\begin{center}
\includegraphics[width=1.0\linewidth]{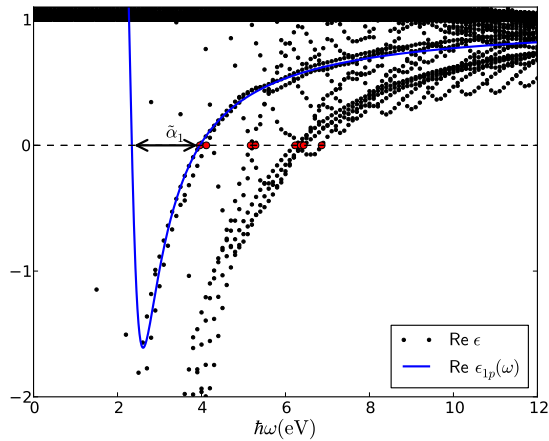}
\caption[system]{\label{fig2} (Color online) Real part of the eigenvalues of the microscopic dielectric function, $\epsilon(\omega)$, for a 10 layer Na film. The frequencies where an eigenvalue crosses the real axis from below (red dots) define the plasmon frequencies, and the corresponding eigenfunctions represent the plasmon modes shown by the red curves in Fig. 1a. The blue curve shows the real part of the single-pole model used to extract the strength of the plasmon. The distance between the zero-points, $\tilde \alpha = \sqrt{\alpha^2-4\gamma^2}$, increases with increasing strength of the mode. }
\end{center}
\end{figure}

In Fig. 1(a) we show the plasmon modes $\phi_n(z)$ (red) and
corresponding charge densities $\rho_n(z)$ (blue) obtained for a 10
atom thick Na film terminated by (100) surfaces. Fig. \ref{fig2} shows the real
part of the eigenvalues of $\epsilon$ for the 10 layer film. The red
dots indicate the plasmon frequencies, $\omega_n$, where the real part
of an eigenvalue crosses the real axis from below. Note that the point
where an eigenvalue crosses the real axis from above corresponds
roughly to the energy of the individual single-particle transitions
contributing to the plasmon state. Thus the distance between the two
crossing points, $\tilde \alpha$, represents the Coulombic
restoring force of the plasma oscillation. The eigenfunctions
of $\epsilon$ corresponding to the indicated eigenvalues are the
plasmon modes shown in Fig. 1. The energies and strengths, $\alpha_n$,
of the plasmon modes are listed to the left. The strength is
determined by fitting a single-pole model
\begin{equation}
 \epsilon_{n,1p} (\omega) = 1-\frac{\alpha_n}{\omega-\omega_{n,0}+i\gamma_n}
\end{equation}
to the value and slope of the relevant eigenvalue branch of $\epsilon$ at the point $\omega=\omega_n$, see Fig. 2.

Returning to Fig.~1, we see that the lowest lying plasmon modes
(S$_1$) are the symmetric and anti-symmetric surface plasmons also
predicted by the classical Drude model. The two sets of modes denoted
S$_2$ and S$_3$ are also localized at the surface, although they
penetrate more into the bulk, and we therefore refer to them as
subsurface modes. The subsurface character of the S$_3$ mode is
perhaps more evident for the 20 layer film where it is more clearly
separated from the bulk modes, see Fig. \ref{fig3}. Experimental loss spectra
of simple metal surfaces have shown a small peak between the surface
and bulk peaks which was assigned to a subsurface plasmon (in that
work denoted multipole surface mode)\cite{17}. Such a peak at
intermediate energies was also observed in a previous RPA calculation
for a jellium surface\cite{26}.  However, the complete analysis of the
plasmons modes presented here shows that more than one subsurface mode
exists.

The bulk plasmons (B) occur at higher energies and resemble standing
waves across the film. The fact that only a discrete set of bulk modes
are observed is clearly a result of confinement of the electron gas
which requires the density to vanish at the boundaries of the film.
The reason only a finite number of modes are found is that the damping
of the modes due to single-particle transitions increases for smaller
oscillation periods, i.e. larger wave number. Consequently, the
strength of the bulk plasmons decreases with increasing wave number
until the point where the real part of $\epsilon$ does not cross the
real axis. This is evident in Fig. 2 where a series of local minima in
the eigenvalue spectrum can be seen at higher frequencies. For all the
films, the highest lying bulk mode has a wave number around 0.5
\AA$^{-1}$. This is very close to the threshold $q$ where
$\text{Re}[\epsilon(q,\omega)]$ becomes positive for all $\omega$ in
bulk Na.

As already mentioned, the main mechanism of energy loss of fast electrons propagating through a material is via excitation of plasmons. In general, the energy dissipated to the electron system due to an applied potential of the form $\phi_{ext}(r,t)=\phi_{ext}(r)\exp(i\omega t)$, is 
\begin{equation}
P(\omega) = \int \int \phi_{ext}(r)\chi_2 (r,r',\omega)\phi_{ext}(r')dr dr'.
\end{equation} 
Here $\chi_2$ is the imaginary part of the density response function, $\chi$. In the case of a fast electron, the external potential is simply that of a point charge moving at constant velocity. We have calculated the loss function for high energy electron beams directed along lines parallel to the film. The resulting EELS spectrum is seen in the middle panel of Fig. 1. It is clear that the loss spectrum is completely dominated by three types of excitations corresponding to the surface, subsurface, and bulk plasmon modes shown to the left. The intensity of the subsurface modes is rather weak in agreement with the low strength ($\alpha$) of these modes.

\begin{figure}
\begin{center}
\includegraphics[width=.7\linewidth]{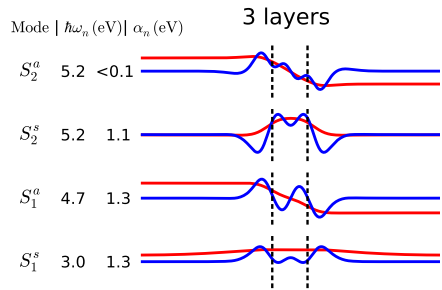}
\caption[system]{\label{fig3a} (Color online) Spatial profile of the plasmon modes of a 3 layer Na films in the direction normal to the film for momentum transfer $q_\parallel$ = 0.1 \AA$^{-1}$. The red (blue) curves show the potential (density) associated with the plasmon excitation, and the mode type, energy and strength ($\alpha$) is shown to the left. }
\end{center}
\end{figure}

\begin{figure}
\begin{center}
\includegraphics[width=.7\linewidth]{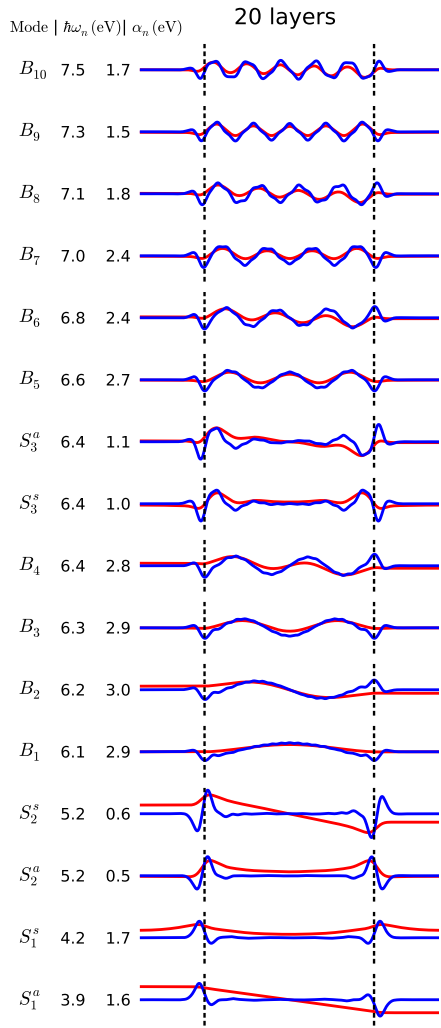}
\caption[system]{\label{fig3} (Color online) Spatial profile of the plasmon modes of a 20 layer Na films in the direction normal to the films for momentum transfer $q_\parallel$ = 0.1 \AA$^{-1}$. The red (blue) curves show the potential (density) associated with the plasmon excitation, and the mode type, energy and strength ($\alpha$) is shown to the left. In addition to the conventional surface modes (S$_1$), the 20 layer slab supports two sets of subsurface plasmon modes (S$_2$,S$_3$) with energy between the surface and bulk modes}
\end{center}
\end{figure}

\begin{figure}
\begin{center}
\includegraphics[width=1.0\linewidth]{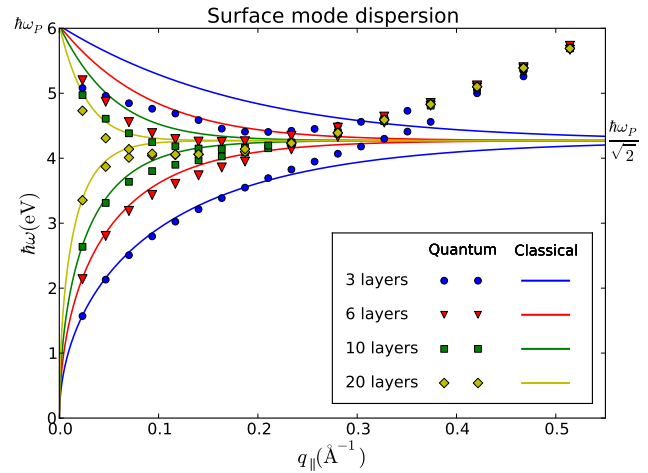}
\caption[system]{\label{fig4} (Color online) Dispersion of surface plasmon energies for different Na film thicknesses. The symbols represent the first-principles RPA results for the energy of the symmetric (lower branch) and anti-symmetric (upper branch) surface plasmons as function of the in-plane wave number $q_\parallel$ . The full lines are the result of a classical Drude model.    }
\end{center}
\end{figure}

Fig. \ref{fig4} shows all the energies of the symmetric and
anti-symmetric surface plasmons (subsurface plasmons not included)
found for four different film thicknesses as a function of the
in-plane wave vector, $q_\parallel$\cite{details_fig4}. The full
curves are the classical Drude results for a 2D metal film with the
electron density of Na. For small $q_\parallel$ the agreement with the
classical result is striking for the symmetric mode. On the other
hand, the quantum results for the anti-symmetric mode are
significantly red shifted compared to the Drude result with deviations
up to 1 eV for the thinnest film. For large $q_\parallel$ the quantum
plasmons show a $q_\parallel^2$ dispersion whereas the classical
result approaches the asymptotic value $\omega_p/\sqrt{2}$. This
failure of the classical model is due to the neglect of the spatial
non-locality of $\epsilon$ , i.e. the $(r-r')$-dependence.

We note that when
exchange-correlation effects are included through the adiabatic local
density approximation (ALDA) kernel, a larger deviation from the
classical model is observed. In particular the anti-symmetric mode is
shifted even further down for small $q_\parallel$. A similar behavior,
i.e. downshift of plasmon energies, is observed for simple metal bulk
systems where the ALDA also predicts a negative dispersion for small
$q$\cite{2}. The spatial form of the plasmon modes obtained with the
ALDA is, however, identical to those obtained at the RPA level.

\begin{figure}
\begin{center}
\includegraphics[width=1.0\linewidth]{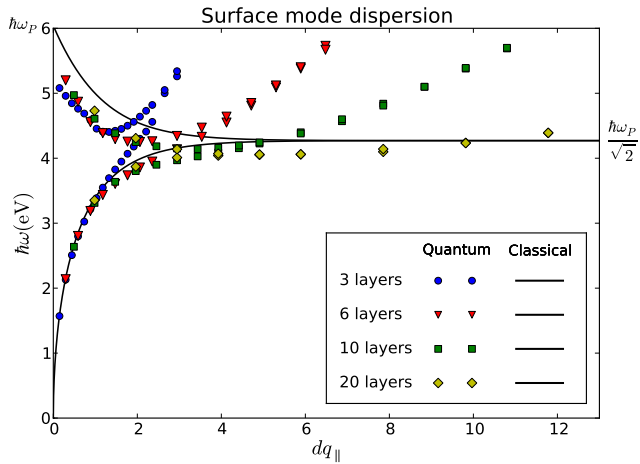}
\caption[system]{\label{fig5} (Color online) Like Fig. \ref{fig4} but plotted as function of the dimensionless parameter $d q_\parallel$ where $d$ is the thickness of the Na film.}
\end{center}
\end{figure}

In Fig. \ref{fig5} we show the same data as shown in Fig. \ref{fig4}
but with the plasmon energy plotted relative to the dimensionless
parameter $d q_\parallel$, where $d$ is the film thickness. When plotted in this way all the classical dispersions fall on the same universal curve. In the regime $d q_\parallel < 2$, it is clear that the quantum results for the anti-symmetric mode are not converged to the classical result even for the thickest slab. 

\begin{figure}
\begin{center}
\includegraphics[width=1.0\linewidth]{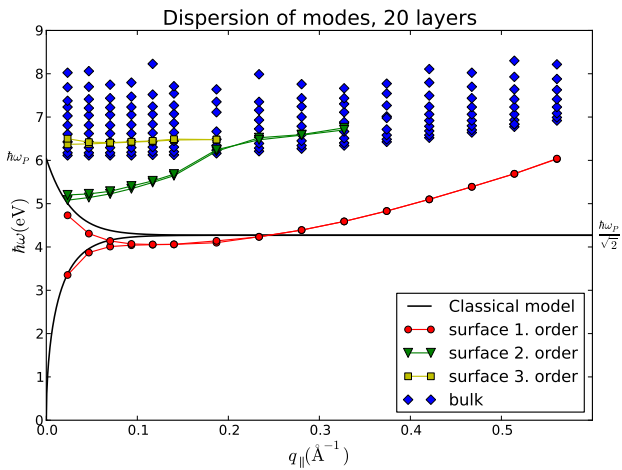}
\caption[system]{\label{fig6} (Color online) Dispersion of all the plasmon modes found for the 20 atomic layer Na film.}
\end{center}
\end{figure}

In Fig. \ref{fig6} we present the dispersion of all plasmon modes
(conventional surface, sub-surface, and bulk modes) for the 20 layer
film. The bulk modes show a weak $q_\parallel^2$ dispersion. The energy offset between the different bulk modes arises from the different wave lenght of the plasmons in the normal direction. The energy of the lowest bulk mode (B$_1$) in the $q_\parallel=0$ limit is $<0.1$ eV higher than the bulk value of $\hbar \omega_p$. This small deviation is due to the finite wave length of the plasmon on the direction perpendicular to the film and this indicates that quantum size effects have a very small influence on the bulk modes of the 20 layer film. 

The first sub-surface modes (green color) follows a
$q_\parallel^2$ dispersion until around $0.2$ \AA$^{-1}$ where it enters
the bulk mode energy range. From this point the dispersion of the
sub-surface mode is reduced and follows that of the bulk modes. The dispersion
of the second sub-surface modes (yellow color) is very similar to the
bulk dispersion. The second sub-surface mode is rather weak and for
$q_\parallel>0.2$ \AA$^{-1}$ its eigenvalue, $\epsilon_n(\omega)$, does
not cross the real axis at any frequency. For both subsurface modes the
splitting between the symmetric and anti-symmetric modes is rather
small. We ascribe this to the weaker strength of the electric field
associated with the subsurface mode compared to the conventional
surface mode: While the charge distribution, $\rho_n$ associated with
the latter has monopole character in the direction perpendicular to
the film, the subsurface modes have dipole character, see Fig. \ref{fig3}.

\section{Conclusion}
We have introduced a method for calculating spatially resolved plasmon
modes in nanostructured materials from first principles.  For the case
of 2D Na films of thickness 0.5-4 nm we found that the
modes could be classified as either surface modes, subsurface modes or
bulk modes. In contrast to previous studies, the direct computation of
eigenmodes revealed that several subsurface modes can exist at the
surface of simple metals. We found clear effects of quantum
confinement on the surface plasmon energies. In particular, the
anti-symmetric surface mode of the thinner films were significantly
red shifted compared to the classical Drude result. Finally, it was
demonstrated how the different features in the calculated spatially
resolved EELS spectrum of the metal films could be unambiguously
ascribed to the excitation of specific plasmon modes.  Apart from
providing an intuitive and visual picture of the collective
excitations of a nanostructure, the spatially resolved plasmon modes
should be useful as a basis for the construction of simple models for
the full non-local dielectric function.

\section{Acknowledgements}
K. S. T. acknowledges support from the Danish Research Council's Sapere Aude Program. The Center for Nanostructured Graphene CNG is sponsored
by the Danish National Research Foundation. Support from The Catalysis for Sustainable Energy (CASE) initiative and the Center of Nanostructuring for Efficient Energy Conversion (CNEEC) at Stanford University, an Energy Frontier Research Center funded by the US Department of Energy, Office of Science, Office of Basic Energy Sciences under Grant no. DE-SC0001060, is also acknowledged.

\appendix
\section{Spectral representation of $\epsilon$}
Let $|\phi_n(\omega)\rangle$ denote the eigenfunctions of the dielectric function,
\begin{equation}
\hat \epsilon(\omega) |\phi_n(\omega)\rangle=\epsilon_n(\omega)|\phi_n(\omega)\rangle
\end{equation}
(we have suppressed the $r$ dependence for notational simplicity). Since $\hat \epsilon(\omega)$ is a non-hermitian operator, the eigenvalues are complex and the eigenfunctions are non-orthogonal. In this case the spectral representation takes the form
\begin{equation}
\hat \epsilon(\omega) = \sum_n \epsilon_n(\omega) |\phi_n(\omega)\rangle \langle \rho_n(\omega)|
\end{equation}
The set $|\rho_n(\omega)\rangle$ is the dual basis of $|\phi_n(\omega)\rangle$ and satisfies
\begin{equation}
\langle \phi_n(\omega)|\rho_n(\omega)\rangle=\delta_{nm}.
\end{equation}
From the spectral representation it follows directly that
\begin{equation}
\hat \epsilon(\omega)^\dagger |\rho_n(\omega)\rangle = \epsilon_n(\omega)^* |\rho_n(\omega)\rangle
\end{equation}
In the following we show that the dielectric eigenfunction $|\phi_n(\omega)\rangle$ and its dual function $|\rho_n(\omega)\rangle$ constitute a potential-density pair, i.e.
\begin{equation}\label{eq.dual2}
\nabla^2 \phi_n(r,\omega)=-4\pi \rho_n(r,\omega)
\end{equation}

Within the RPA, the dielectric function is related to the non-interacting polarization function, $\chi_0(r,r',\omega)$, by 
\begin{equation}
\hat \epsilon(\omega)=\hat 1-\hat v \hat \chi_0(\omega)
\end{equation} 
where $\hat v=1/|r-r'|$ is the Coulomb interaction, $\hat 1= \delta(r-r')$. Exploiting that, under time reversal symmetry $\chi_0(r,r')=\chi_0(r',r)$, we have
\begin{equation}
\hat \epsilon(\omega)^\dagger =\hat 1-\hat \chi_0(\omega)^* \hat v
\end{equation}
We now make the ansatz $\rho_n(\omega)=\hat v^{-1}\phi_n(\omega)^*$ and evaluate
\begin{eqnarray}
\epsilon(\omega)^\dagger \hat v^{-1}\phi_n(\omega)^*&=&  v^{-1}\phi_n(\omega)^*- \hat \chi_0(\omega)^*\phi_n(\omega)^*\\
&=&\hat v^{-1}[\hat \epsilon(\omega) \phi_n(\omega)]^*\\
&=&\epsilon_n(\omega)^*\hat v^{-1}\phi_n^*
\end{eqnarray}
which concludes the proof.



\begin{thebibliography}{31}
\bibitem{note}
In the case of Ag films (results will be published elsewhere) we have found that the coupling to interband transitions can shift the peak of $\text{Im}\epsilon_n(\omega)$ away from the point where $\text{Re}\epsilon_n(\omega)=0$. Importantly, however, the corresponding eigen function $\phi_n(r,\omega)$ do not change significantly when $\omega$ is varied between the frequencies given by the two criteria and thus the spatial form of the plasmon remains well defined.
\bibitem{details_fig4}
To obtain the dispersion of the surface plasmons (Fig. 4) of an isolated film, the calculated plasmon energies were extrapolated to the limit of infinite distance between the repeated films using a classical model for the plasmon energies in an infinite array of films. The model is a straightforward generalization of the results in Ref. \cite{27}
\end{thebibliography}
\end{document}